Maria RAMARD[1,2*],
Romain LANIEL[2],
Mathieu MIROIR[2],
Olivier KERBRAT[1,2]

# QUANTIFICATION OF THE INFLUENCE OF MORPHOLOGIES ON LASER CUTTING QUALITY

## 1. INTRODUCTION

Laser cutting has existed since 1967, when Peter Houldcroft used an oxygen-assisted gas system to cut a 1-mm-thick-steel sheet with a $CO_2$ laser beam. This initiative was launched by the American aerospace industry, which concluded that laser cutting could be an efficient and cost-effective cutting tool [1].

From the 1990s to 2020s, several studies have been conducted to understand the mechanisms involved in heating metals and to improve cut quality. Figure 1 illustrates the laser cutting process and the main parameters that have been studied in the literature.

Early studies primarily focus on minimising a specific defect such as the Heat-Affected Zone [2, 3], surface roughness [4, 5] or kerf geometry [6, 7]. Most of these studies focus on specific types of cutting defects and improve quality by testing different parameters. However, these studies aim to minimise specific defects rather than employing a comprehensive methodology that considers all cutting defects. This trend is due to the rapid industrialisation, which has given priority to practical optimisation over a deeper understanding of the underlying physics. Research has largely focused on optimising laser manufacturing parameters to improve quality, as illustrated in Fig. 1 (cutting speed, fibre source power, laser beam pulse frequency, assist gas pressure, focus point, nozzle diameter). Several studies have proposed parameter settings to minimise defects, mainly by varying the speed, power and pressure of the assist gas. In addition, some studies have looked at the influence of part parameters, such as material properties, thickness and, above all, morphologies, which influence cutting trajectories. Although they have received less attention, Fig. 1 shows the two morphologies that have been studied to improve quality as drill hole and angle size.

___________

[1] Mécatronique, ENS Rennes, France
[2] Mécanique et Verres, Institut de Physique de Rennes, France
* E-mail: maria.ramard@ens-rennes.fr
   https://doi.org/10.36897/jme/203192

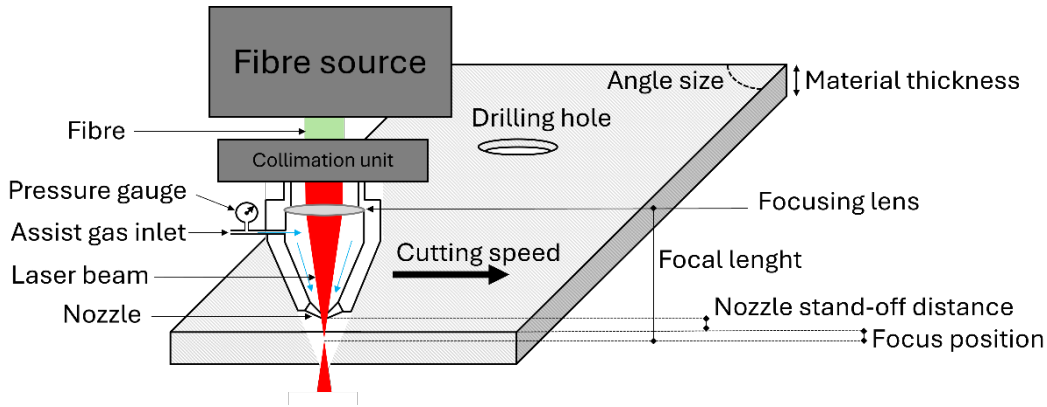

Fig. 1. Fibre laser cutting diagram

The first aim of this study is therefore to demonstrate that, as well as manufacturing parameters, part parameters, and in particular morphologies, have a significant impact on the appearance of all laser cutting defects. The second aim is to show that it is possible to identify design limitations and to help designers with their design choices.

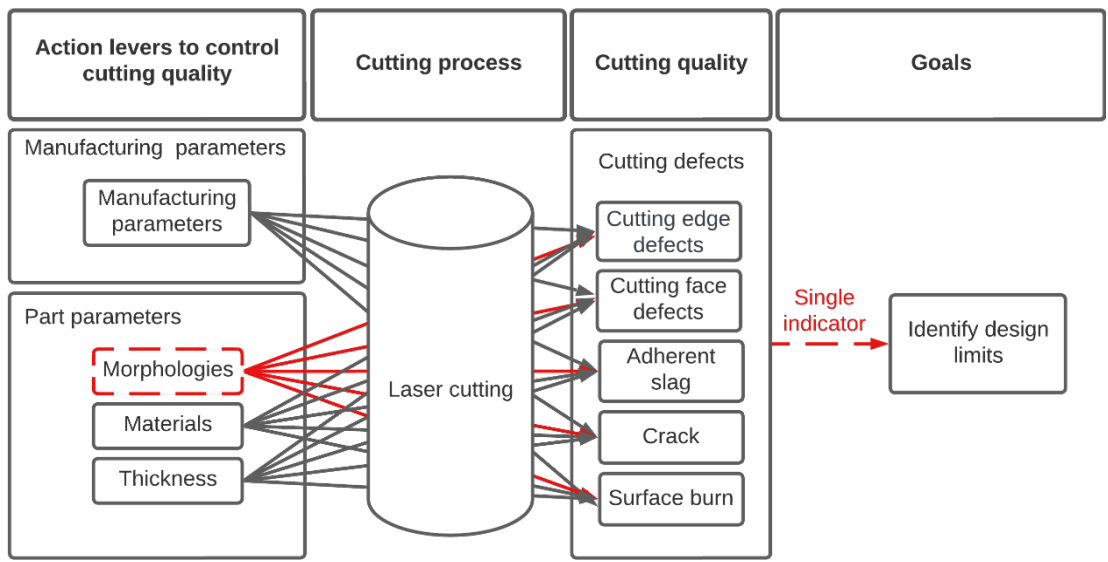

Fig. 2. Aims of this study, shown in red

Figure 2 illustrates the aims of this study. Firstly, this study proposes to use a method that enables a single indicator to be defined for comparing the different cutting defects of the same part, regardless of the nature and number of its morphologies. This method makes it possible to integrate the morphologies into the optimisation parameters in order to demonstrate their impact on the quality of cut. In addition, this single indicator enables all cutting defects to be compared and ranked in order to identify design limits. For example, a thickness limit could be identified for each cutting defect for a given material and morphology, which would facilitate optimisation choices. This is the second aim of this study.

## 2. STATE OF THE ART

To understand the scientific community's motivation for optimising the quality of laser cutting, it is important to remember that this process is a non-contact machining technology. The laser beam, acting as a heat source, is focused on a very small area of the workpiece. This multi-physical process causes significant phase changes within the material. This leads to thermal effects for metal cut parts that cause various types of known and standardised cutting defects, such as burrs, irregular surface profiles, geometric deviations, adherent slag, cracks visible to the naked eye and surface burns (EN ISO 17658).

The parameters selected in the literature to determine cutting quality include both optimisation parameters and non-variable parameters. The latter are well studied and known, such as the material. Its nature is a determining parameter for quality, but it cannot be used as a lever to improve it. Therefore, before cutting, it's essential to understand the physico-chemical properties of the material to be cut. These properties have been identified in several literature reviews focusing on the laser cutting of these materials. An earlier review by Avanish Kumar Dubey in 2008 [8] concluded that the performance of laser beam machining mainly depends on laser parameters (e.g. laser power, wavelength, modes of operation), material parameters (e.g. type, thickness) and process parameters (e.g. feed rate, focal plane position, frequency, energy, pulse duration, assist gas types and pressure) in the same way as two more recent reviews of Amit Sharma and Vinod Yadava in 2017 [9] and Pushkal Badoniya in 2018 [10]. Based on these findings, it can be concluded from the literature that the key optimisation parameters include laser parameters, material thickness and process parameters.

However, concerning morphologies, some studies have also examined their influence. Indeed, although most studies carried out experimental designs by cutting segments of different lengths [11–14] without considering the influence of morphology on quality, two articles highlight the importance of taking morphology into account. Firstly, the angle size was studied as a significant parameter affecting cutting quality. The study found that the sharper the angle, the more excess melt accumulated in the corners [15]. Secondly, another study proposed a predictive visualisation of the diameter and height of a laser-cut hole using a neural network model [16]. These two studies consider morphology as an influent parameter to better understand the quality of the process. An analysis of the literature reveals that part parameters are less frequently considered in quality optimisation compared to manufacturing parameters, as shown in Table 1. Material thicknesses are included in the category of part parameters, and it is possible to identify two studies looking at their influence, one by Wandera C. et al [17] and the other by Teixidor D. et al [18].

Table 1 lists the parameters used in studies aimed at improving the quality of cut. Standardised cutting defects have been listed and it can be seen that adherent slag and cracks have been less studied. Adherent slag is not studied, or at least this terminology is not used in the literature. The focus is on excess melt or burrs. Cracks are not studied for quality optimisation; they're a consequence of surface burn, known in the literature, but which the laser cutting research community is not trying to reduce. Indeed, the study of Xiao-Nan Wanga, Qian Sunb, Zhi Zhenga and Hong-Shuang Di in 2017 explained that previous studies have shown that heating the material leads to a decrease in hardness and causes cracks to appear. These cracks are a consequence of the HAZ where the hardness of the material decreases [19].

Table 1. List of manufacturing and part parameters used in studies to improve the cutting quality of laser-cut parts

| Standardised cutting defects by EN ISO 17658 | Manufacturing parameters | Part parameters |
|---|---|---|
| Burr | Cutting speed [15, 17]<br>Focus point [17]<br>Assist gas pressure [15, 17]<br>Nozzle diameter [17] | Angle size [15]<br>Drilling hole [16] |
| Irregular cutting face profile | Power [2, 3, 11]<br>Cutting speed [2, 3, 4, 5, 9, 11, 13, 15]<br>Assist gas pressure [2,4,5,9,11,15]<br>Frequency [9]<br>Focus point [5,13] | Material thickness [18] |
| Geometric deviation | Power [6, 7, 12]<br>Cutting speed [6, 7, 12]<br>Assist gas pressure [6–7]<br>Frequency [6] | |
| Adherent slag | | |
| Crack | | |
| Surface burn | Power [2, 3, 11]<br>Cutting speed [2, 3, 11]<br>Assist gas pressure [2, 11]<br>Frequency [2] | Material thickness [17] |

The goal of this paper is to propose a more comprehensive framework for improving cutting quality than what is currently available in the literature. To characterise and compare the influence of each part parameter on each cutting defect, a common indicator must be identified. The study, therefore, proposes using the FMECA (Failure Modes, Effects and Criticality Analysis) method, substituting cutting defects for machine failures. The criticality calculated in this way will be used to assess the quality of the cut parts.

Indeed, criticality can be determined using various methods, including FMECA (Failure Modes, Effects and Criticality Analysis), which has been standardised in an IEC 60812:2018 and is widely used by the industrial sector. This method is often integrated into continuous improvement projects for production lines. It's used to qualify and quantify machine failures to assess their criticality. The results are used to prioritise optimisation actions, targeting the most critical machines to ensure the reliability and availability of production resources, reduce downtime and improve maintenance. This is a FMECA for production resources. A recent study used this method to identify the faultiest production line and the bearings of the most critical machines [20]. An earlier study used this method to identify improvements to be made to the design and maintenance of a mechanical system. This was a FMECA of a product [21]. This method seems well suited to our needs. It allows different types of failure to be compared. It also makes it possible to quantify them using a common indicator - criticality - whatever the systems studied.

First, the method will be reviewed. Then, a case study will be presented to evaluate the quality of laser-cut metal parts.

Finally, the results will allow us to conclude on the influence of part parameters, particularly morphologies, on all types of lasers cutting defects.

## 3. METHOD

The first step of FMECA method is to define the criticality factor matrix $C$, consisting of terms $C_{ij}$ where cutting defects are identified by $j$ and morphologies identified by $i$.

First, a batch of $N$ identical parts must be selected. Secondly, each criticality factor is evaluated, which requires the definition of the range of values for each bounded factor beforehand. The IEC 60812:2018 FMECA standard recommends adapting these ranges to the specific context of the study to ensure consistency and accuracy. The number of levels used for each factor should be chosen according to the nature of the application, as scales that are too granular or insufficiently detailed can affect the relevance of the assessment. This approach ensures that the ranges selected remain meaningful and correctly reflect the criticality factors within the scope of the study. The detectability of a defect within a morphology $D_{ij}$ requires an upstream detection method and is considered as a binary factor belonging to $\{D_{min}; D_{max}\}$. The probability of occurrence $O_{ij}$ is derived from the probability of the defect occurring on the morphology $P_{ij}$, calculated as the ratio of the number of morphologies affected by the defect to the total number of morphologies in the batch. The parameters $O_{ij}$ are determined by a linear relation of $P_{ij}$ and bounded by two limits $O_{min}$ and $O_{max}$.

Then, the gravity of the defect in the morphology $G_{ij}$ is determined as the sum of identified effects. These effects are assessed according to three levels of significance:
- If very significant, the elementary effect value equals the ratio of $G_{max}$ to the number of effects;
- If moderately significant, it equals $G_{avg}$ divided by the number of effects;
- If not significant, it equals to $G_{min}$ divided by the number of effects.

Clear definitions of gravity effects and the associated measurement methods are crucial. It is also important to ensure the independence of the factors when defining them. Finally, criticality corresponds to the Hadamard matrix product (1).

$$\boldsymbol{C = D \circ O \circ G} \qquad (1)$$

## 4. CASE STUDY

The method is tested on a test-part application. The selected process for this study is a laser-cutting machine, focusing on the cutting of batches containing 10 identical parts made from DC01 steel (C: 0.12%, Mn: 0.6%, P: 0.045%, S: 0.045%). The thicknesses examined range from 0.6 mm to 4 mm, specifically {0.6, 0.8, 1, 1.2, 1.5, 2, 3, 4} mm. The machine has a 3000 W fibre source and nitrogen as the assist gas. Manufacturing parameters vary depending on the thickness, based on current production settings.

The selected parts have been specially designed to test all possible laser-cutting morphologies, which will be defined in the following section. Each batch will be evaluated to quantify the criticality of the various cutting defects on the morphologies in accordance with the proposed method.

### 4.1. DEFINITION OF MORPHOLOGIES

To test all possible laser-cutting morphologies, three types were defined: arcs, segments and angles. These primary morphologies are defined thanks to a review of the literature on morphologies cut out in quality optimisation studies and a successful exercise which consisted of completing all the parts of an industrial catalogue solely with these three categories. These primary morphologies are further classified based on characteristic dimensions that define secondary morphologies, allowing for more precise analysis. In the industrial application of this study, this classification is based on distinctive criteria established with input of design and production experts (Table 2).

Table 2. Definition of morphologies with a colour assigned for better mapping on a part

| Morphologies | Definition | Morphology class names | Class $i$ | Discriminating ranges of characteristic dimensions |
|---|---|---|---|---|
| Arc | Arcs are defined by a centre O, a radius R and an angle α. A circle is a closed plane curve with centre O and radius R where α=360°. | Small arc | 1 | [0;16] mm |
| Arc | | Long arc | 2 | ]16;60] mm |
| Segment | A segment is a set of points aligned between two points A and B and defined by a length. | Segment | 3 | [0;2000] mm |
| Angle | An angle is formed by two intersecting straight lines. | Acute angle | 4 | ]0;90 [° |
| Angle | | Right and obtuse angle with an easier rework | 5 | [90;180[U]180;360 [° |
| Angle | | Right and obtuse angle with a more difficult rework | 6 | [90;180[U]180;360 [° |

Classes 1 to 3 correspond to the radius and segment dimensions most used for the test company's industrial parts. For arcs, morphology classes are distinguished solely by radius. The angle of the arc is not a discriminating factor in this industrial case. For segments, only one class is defined, since, according to production experts, the dimension of this morphology does not significantly affect variations in cutting defects. Classes 4 to 6 correspond to angles, which are classified as acute angles between ]0;90 [° and right and obtuse angles ranging from [90;180 [U]180;360 [°. Flat angles are not considered. The classification of angles was determined with expert input, based on the difficulty of reworking defects. Acute angles are identified as the most difficult to rework, which aligns with findings in the literature [15]. Then, according to the experts, some right and obtuse angles are more difficult to rectify than others. A classification was therefore proposed for these angles, which explains why there are two morphologies identified for these angles.

Table 2 defines the six morphologies identified for this study, with each morphology assigned a specific colour for easy mapping. This detailed classification makes it clear that a part can be composed of different morphologies, each present in varying quantities.

Figure 2 represents the selected part of this study, including all morphologies.

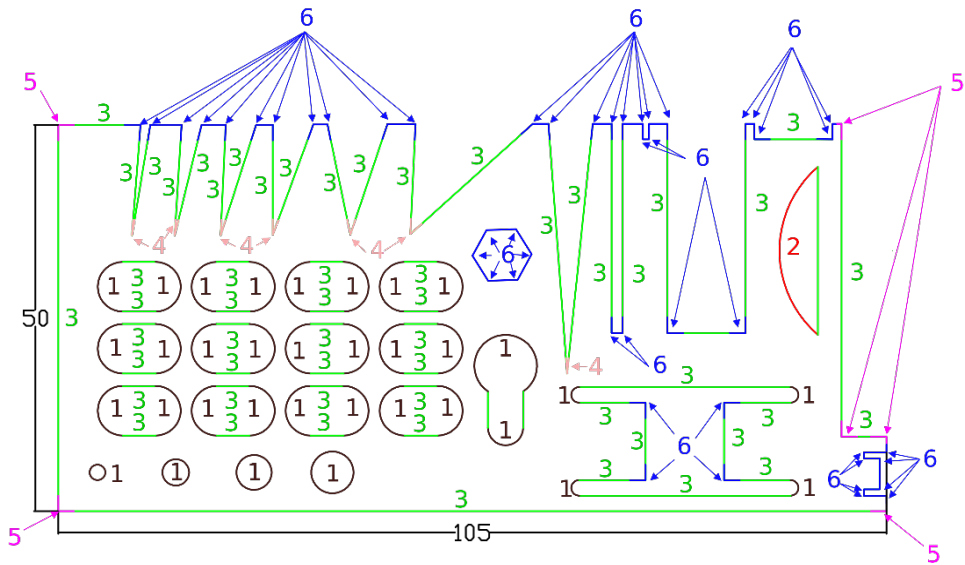

Fig. 3. Test part, with mapping of with mapping corresponding to the definition of associated morphologies (Table 2)

These last two figures and tables can be used to map any part and quantify the presence and number of all the morphologies in order to determine the potential risk of quality problems to the naked eye.

### 4.2. EVALUATION OF THE CRITICALITY OF TESTED PARTS

The following laser cutting defects are indexed from 1 to 6 respectively: burrs, irregular surface profiles, geometric deviations, adherent slag, crack visible to the naked eye and surface burns. Morphologies are also defined and indexed from 1 to 6.

In this application, detectability is assessed on a scale from 1 to 2, the probability of occurrence on a scale from 1 to 3, and gravity on a scale from 1 to 4. While the conventional FMECA approach often uses a 1-to-10 scale, the standard does not impose a fixed numerical range but rather allows for adaptation based on the context and objectives of the analysis. Here, the scales are limited to 4 because the factors are determined through calculation. The chosen scales align with the nature of our analysis and the constraints of our industrial application. Expanding these scales further would not provide additional meaningful discrimination. Detectability, being a binary factor, is fully represented by a scale from 1 to

2. The probability of occurrence is defined using parameter $P_{ij}$, which classifies the evaluated parts into three categories: unlikely, moderately likely, and highly likely. To give greater weight to gravity while maintaining mathematical consistency among the factors, the gravity scale is set between 1 and 4, resulting in criticality ranging from 1 to 24. The probability of occurrence is considered more significant than detectability because it determines whether a defect is more likely to appear in one morphology than another. Lastly, gravity is regarded as the most important factor, as a severe defect can lead to multiple consequences.

Once the minimum and maximum for each factor are established, it is important to evaluate them.

First, a detection method must be defined. In this study, the following approach is used: if a defect is easy to see with the naked eye at the end of production, then $D_{ij} = D_{min}$. If it needs a more precise detection method, then $D_{ij} = D_{max}$. A detectability matrix $\boldsymbol{D}$ is then created for each defect type and morphology (2).

$$\boldsymbol{D} = \begin{bmatrix} 1 & 1 & 1 & 1 & 2 & 1 \\ 1 & 1 & 1 & 1 & 2 & 1 \\ 1 & 1 & 1 & 1 & 2 & 1 \\ 1 & 2 & 1 & 1 & 2 & 1 \\ 1 & 2 & 1 & 1 & 2 & 1 \\ 1 & 2 & 1 & 1 & 2 & 1 \end{bmatrix} \tag{2}$$

In this application, cracks are harder to detect and need a more precise detection method, then $D_{i5} = 2$. In addition, irregular cut face profiles are easier to observe on arcs and segments because these morphologies have continuous shapes. However, it's harder to detect on angles due to the sudden changes in the profile's shape, which makes detection more difficult. Therefore, $D_{42} = 2, D_{52} = 2$ et $D_{62} = 2$.

Secondly, to assess the probability of occurrence of the defect, it is necessary to count the number of morphologies that exhibit the defect and divide it by the total number of morphologies in the batch of parts studied. The probability of occurrence is defined from $O_{min} = 1$ to $O_{max} = 3$.

Thirdly, the effects associated with gravity need to be assessed. To evaluate the criticality of cutting defects, this study proposes to define only one gravity effect for this industrial case like the difficulty of recovering from a defect in a morphology.

In this case study, the operator's experience was used to classify the morphologies by the level of recovery difficulty bounded by $G_{min} = 1$ and $G_{max} = 4$.

A gravity matrix $\boldsymbol{G}$ is then created for each defect type and morphology (3).

$$\boldsymbol{G} = \begin{bmatrix} 4 & 4 & 4 & 4 & 1 & 1 \\ 2.5 & 2.5 & 4 & 2.5 & 1 & 1 \\ 1 & 1 & 4 & 1 & 1 & 1 \\ 4 & 4 & 4 & 4 & 1 & 1 \\ 1 & 4 & 4 & 1 & 1 & 1 \\ 2.5 & 4 & 4 & 2.5 & 1 & 1 \end{bmatrix} \tag{3}$$

Geometric deviations ($j = 3$) have the highest gravity because this defect cannot be reworked and systematically generates scraps. Secondly, cracks ($j = 5$) and surface burns ($j = 6$) are not reworked in this industrial case study because they are not severe defects that require reworking. The parts are painted after this process, so this is not a significant defect at this scale. Other defects need to be assessed beforehand by the operators to establish a gravity matrix (3).

Once the morphologies, defects, effects and scales of each factor have been defined, criticality can be calculated. This study includes the analysis of 80 parts, resulting in 288 criticality assessments calculated for each defect and morphology present in each batch of varying thicknesses.

## 5. RESULTS

The results of this study are presented in Fig. 3, illustrating the criticality of laser cutting defects as a function of part parameters, specifically the thicknesses and morphologies of the test parts. The criticality is represented from 1 to 12, the maximum criticality calculated for these test parts.

Figure 3a depicts the evolution of the criticality of burrs in relation to thickness and different morphologies. Burrs are more critical for certain morphologies such as arcs, acute angles and right and obtuse angles with a more difficult rework. Criticality varies with thickness, allowing for the identification of a limit thickness beyond which the criticality of burrs increases.

In contrast, the evolution of the criticality for burrs on segments and for right and obtuse angles with an easier rework is much less significant. This indicates that all morphologies influence burrs, but the criticality depends on their nature and the thickness of the material.

Figure 3b illustrates the criticality of irregular cutting face profiles and highlights a different evolution compared to burrs. The criticality of this defect does not change with the thickness; it is a defect that is systematically present, with criticality depending solely on morphology. Arcs and angles appear to be the most sensitive morphologies for this defect.

Figure 3c shares the same characteristics as Figure 3b, as there is no variation in criticality as a function of thickness, but no variation in morphology either. In fact, this defect is very rare, and the detectability and gravity are similar for all morphologies. In these experimental parts, no geometric deviation was observed, and the criticality is therefore equal to 4 for all morphologies and thicknesses (1).

Like the criticality of burrs illustrated in Fig. 3a, some morphologies exhibit greater criticality for adherent slag than others, as illustrated in Fig. 3d. In decreasing order, the morphologies with the most critical adherent slag are acute angles, small arcs, long arcs, right and obtuse angles with an easier rework, right and obtuse angles with a more difficult rework and finally segments.

Figure 3e depicts the criticality of cracks. Like the criticality of geometric deviations illustrated in Fig. 3c, the criticality does not vary as a function of thickness or morphology. This is because cracks are not detected by a more precise method in this application and are not significant for the quality of the parts that will be painted after laser cutting.

Lastly, the criticality of surface burns, as shown in Fig. 3f, is almost constant for all thicknesses. The criticality is highest for acute angles.

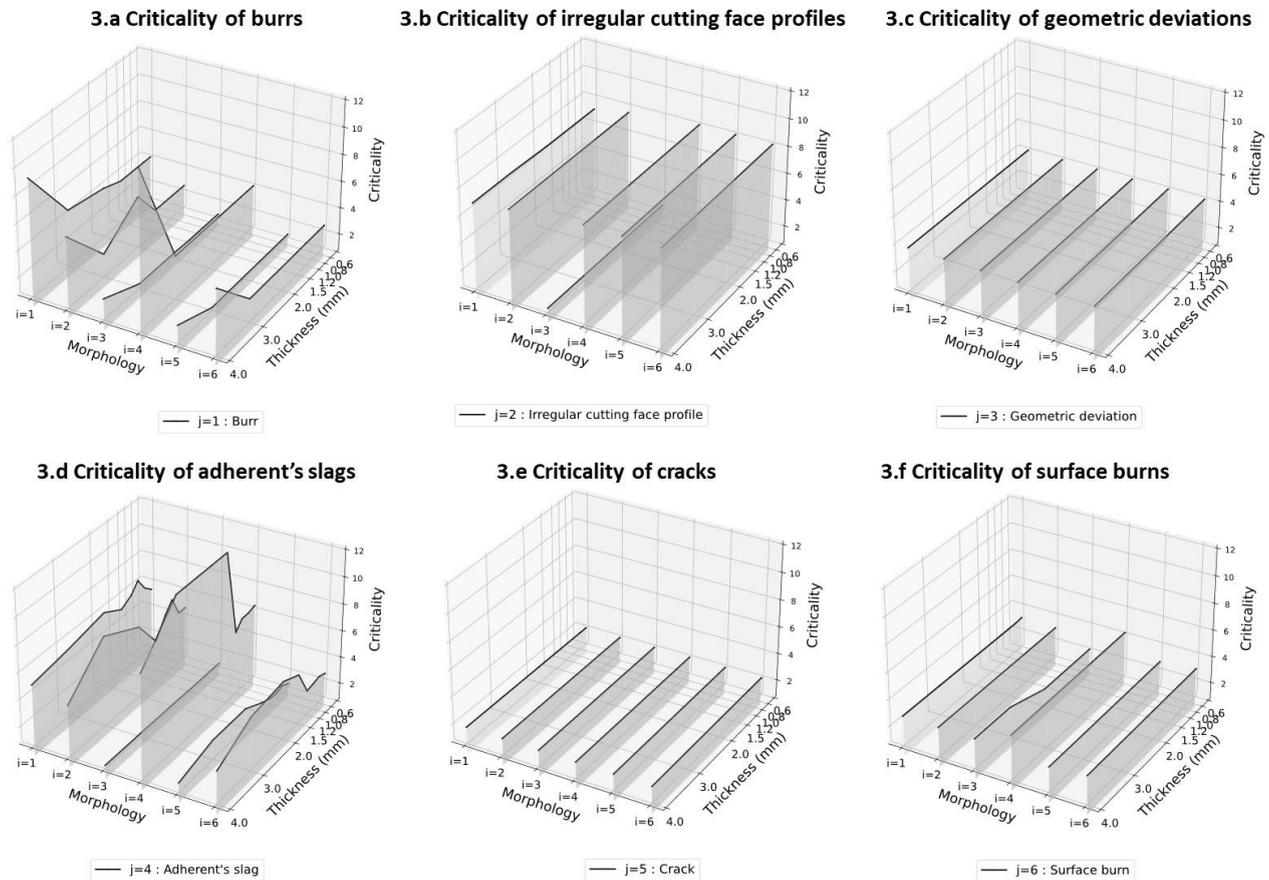

Fig. 4. Criticality of cutting defects on morphologies (Table 2) of test parts cut from DC01 steel

The analysis of these results allows for ranking the defects from the most critical of the least critical for each morphology. Burrs (Figure 3.a), irregular surface profiles (Fig. 3b) and adherent slag (Fig. 3d) are the most critical cutting defects in these batches of parts, particularly on arcs and angles. These defects are directly linked to the heating of the material by the laser energy input.

Moreover, among these defects, only burrs and adherent slag are influenced by the thickness, so the study highlights that it is possible to identify a focus on minimising these defects. A key point to note from these results is that the morphology of the part influences all the defects.

## 6. DISCUSSION

The results have identified the most critical defects for each morphology and thickness limit. These results open the way to the definition of design limits, as shown in Table 3.

Table 3. Design limits identified with criticalities of test parts (Fig. 3)

| Defects / Thickness / Morphologies | j=1 | | | | | | | | j=2 | | | | | | | | j=3 | | | | | | | | j=4 | | | | | | | | j=5 | | | | | | | | j=6 | | | | | | | |
|---|---|---|---|---|---|---|---|---|---|---|---|---|---|---|---|---|---|---|---|---|---|---|---|---|---|---|---|---|---|---|---|---|---|---|---|---|---|---|---|---|---|---|---|---|---|---|---|---|
|  | 0.6 | 0.8 | 1 | 1.2 | 1.5 | 2 | 3 | 4 | 0.6 | 0.8 | 1 | 1.2 | 1.5 | 2 | 3 | 4 | 0.6 | 0.8 | 1 | 1.2 | 1.5 | 2 | 3 | 4 | 0.6 | 0.8 | 1 | 1.2 | 1.5 | 2 | 3 | 4 | 0.6 | 0.8 | 1 | 1.2 | 1.5 | 2 | 3 | 4 | 0.6 | 0.8 | 1 | 1.2 | 1.5 | 2 | 3 | 4 |
| i=1 | | | | | | | | | | | | | | | | | | | | | | | | | | | | | | | | | | | | | | | | | | | | | | | | |
| i=2 | | | | | | | | | | | | | | | | | | | | | | | | | | | | | | | | | | | | | | | | | | | | | | | | |
| i=3 | | | | | | | | | | | | | | | | | | | | | | | | | | | | | | | | | | | | | | | | | | | | | | | | |
| i=4 | | | | | | | | | | | | | | | | | | | | | | | | | | | | | | | | | | | | | | | | | | | | | | | | |
| i=5 | | | | | | | | | | | | | | | | | | | | | | | | | | | | | | | | | | | | | | | | | | | | | | | | |
| i=6 | | | | | | | | | | | | | | | | | | | | | | | | | | | | | | | | | | | | | | | | | | | | | | | | |

Legend: ▇ $C<4$    ▇ $4 \leq C < 8$    ▇ $C \geq 8$

Table 3 shows the design limits using the established threshold. For this application, the maximum criticality calculated is equal to 12, so it is possible to colour in green the criticalities of defects on the morphology that are less than 4. Then, in yellow, those between 4 and 8. And in red, those above 8. This analysis gives a better view of the design limits and provides designers with an initial design support tool. For small arcs, adherent slag is more critical within the 0.8 mm to 3 mm thickness range. However, from 3 mm to 4 mm in thickness, the trend reverses, with burrs becoming twice as critical as adherent slag. Similarly, for long arcs, the evolution of the criticality of adherent slag and burrs varies with thickness. A decrease in criticality is observed from 2 mm upwards (Fig. 3), and this decrease is more significant for adherent slag than for burrs when compared with the previous morphology (Table 3). This indicates that as the arc diameter increases, the criticality of the adherent slag decreases beyond a certain thickness. These results suggest possible thickness and morphology limits, particularly in terms of diameter.

Segments are the least sensitive to the appearance of critical cutting defects. This morphology is widely used in the literature of studies utilising the design of experiments to improve cut quality. However, the influence of morphology is not considered, and these results underline the fact that the tests are carried out on the morphology that is least sensitive to the appearance of critical defects. This finding confirms the lack of exhaustiveness of existing studies [11–14] and underscores the need to study the influence of morphologies to improve quality.

As Table 2 shows, acute angles are the most sensitive to the formation of burrs as soon as the thickness exceeds 3 mm and adherent slag as soon as the thickness exceeds 1.2 mm. Therefore, it is essential to focus first on the quality of acute angles. This observation aligns with the results of the study of Dong-Gyu Ahn and Young-Tae Yoo (2006), which demonstrates that the size of the angle influences the appearance of excess melt. It indicates that the sharper the angle, the more sensitive it is to the occurrence of cutting defects [15]. This study clarifies this influence by demonstrating that it is necessary to minimise burrs of 3 mm to 4 mm and adherent slag of 0.6 mm and 1.2 mm and above all of 1.5 mm to 4 mm thickness.

These results allow a ranking of defects in terms of quality and quantity for each design input. Indeed, the analysis revealed that the most critical defects are thermal, especially burrs and adherent slags. Furthermore, the results highlight that the morphologies most sensitive to these defects are arcs and angles, especially beyond a certain thickness and characteristic dimension.

This analysis could be used to initiate research into the optimisation of laser cutting quality in a comprehensive and justified manner.

## 7. CONCLUSION

First, this study demonstrates that part parameters, particularly morphologies, significantly impact cutting quality. Three types of morphology were defined in this study to characterise any laser-cut part: arcs, segments and angles. Their influence was quantified on the six types of defects defined in standard EN ISO 17658, namely: burrs, irregular surface profiles, geometric deviations, adherent slags, cracks and burns. This study fills a gap in the literature by showing how important it is to take morphologies into account when choosing parameters to optimise the quality of the cut, with the results showing that certain morphologies are more influential than others, depending on the nature of the defect.

Then, identifying a common indicator, such as criticality, allows for a global comparison of part quality. This also helps to identify design limits and gives designers useful tools. For example, in the industrial case of this study, the method used made it possible to identify the two most influential morphologies on two types of defects, namely arcs and angles, which influence the appearance of burrs and slag, which are very critical for quality.

Integrating these results is essential to improve quality effectively. It may be possible to combine this analysis of defect criticality with manufacturing parameters for multi-criteria optimisation. Then, some preliminary tests have been carried out on parts similar to this study, cut from stainless steel using the same process. Other similar parts were also manufactured using a flame-cutting process, as the EN ISO 17658 standard is the same for cutting defects in flame-cutting, laser-cutting and plasma-cutting. The results showed that it was possible to prioritise cutting defects according to design parameter using other materials and other thermal cutting processes.

Finally, it would be useful to carry out a more comprehensive study in order to propose recommendations for adapting this method to different types of parts in order to study other materials, other thicknesses and other processes. This study demonstrates that it is possible to provide information on the current quality level of manufactured parts and analysis tools for subsequent quality optimisation.

## REFERENCES

[1] HILTON P.A., 1997, *Early Days of Laser Cutting*, Lasers in Material Processing, 3097, 10–16, https://doi.org/10.1117/12.281076.
[2] PATEL A., BHAVSAR S.N., 2020, *Experimental Investigation to Optimize Laser Cutting Process Parameters for Difficult to Cut Die Alloy Steel Using Response Surface Methodology*, Materials Today: Proceedings, 43, 28–35, https://doi.org/10.1016/j.matpr.2020.11.201 2214-7853.
[3] CAYDAŞ U., HASCALIK A., 2008, *Use of the Grey Relational Analysis to Determine Optimum Laser Cutting Parameters with Multi-Performance Characteristics*, Optics and Laser Technology, 40/7, 987–994, https://doi.org/10.1016/j.optlastec.2008.01.004.
[4] ZHANG Y.L., LEI J.H., 2017, *Prediction of Laser Cutting Roughness in Intelligent Manufacturing Mode Based on ANFIS*, Procedia Engineering, 174, 82–89, https://doi.org/10.1016/j.proeng.2017.01.152.
[5] GARCIA A.T., LEVICHEV N., VORKOV V., RODRIGUES G.C., CATTRYSSE D., DUFLOU J.R., 2020, *Roughness Prediction of Laser Cut Edges by Image Processing and Artificial Neural Networks*, Procedia Manufacturing, 54, 257–262, https://doi.org/10.1016/j.promfg.2021.07.040.

[6]   NORKEY G., SINGH K.P., PRAJAPATI A., SHARMA V., 2019, *Intelligent Parameters Optimization for Laser Cutting of Highly Reflective and Thermally Conductive Materials Using Artificial Neural Network*, Materials Today: Proceedings, 46, 4757–4764, https://doi.org/10.1016/j.matpr.2020.10.309.

[7]   KARTHIKEYAN R., SENTHILKUMAR V., THILAK M., NAGADEEPAN A., 2018, *Application of Grey Relational Analysis for Optimization of Kerf Quality During $CO_2$ Laser Cutting of Mild Steel*, Materials Today: Proceedings, 5, https://doi.org/10.1016/j.matpr.2018.06.276.

[8]   DUBEY A.K., YADAVA V., 2008, *Laser Beam Machining-a Review*, International Journal of Machine Tools and Manufacture, 48/6, 609–628, https://doi.org/10.1016/j.ijmachtools.2007.10.017.

[9]   SHARMA A., YADAVA V., 2018, *Experimental Analysis of Nd-YAG Laser Cutting of Sheet Materials–a Review*, Optics & Laser Technology, 98, 264–280, https://doi.org/10.1016/j.optlastec.2017.08.002.

[10]  BADONIYA P., 2018, *$CO_2$ Laser Cutting of Different Materials—a Review*, Int. Res. J. Eng. Technol., 5/6, 1–12, https://www.irjet.net/volume5-issue06, Accessed 9 July 2024.

[11]  PATEL P., SHETH S., PATEL T., 2016, *Experimental Analysis and ANN Modelling of HAZ in Laser Cutting of Glass Fibre Reinforced Plastic Composites*, Procedia Technology, 23, 406–413, https://doi.org/10.1016/j.protcy.2016.03.044.

[12]  YILBAS B.S., SHAUKAT M.M., ASHRAF F., 2017, *Laser Cutting of Various Materials: Kerf Width Size Analysis and Life Cycle Assessment of Cutting Process*, Optics and Laser Technology, 93, 67–73, https://doi.org/10.1016/j.optlastec.2017.02.014.

[13]  WANDERA C., SALMINEN A., OLSEN F.O., KUJANPÄÄ V., 2006, *Cutting of Stainless Steel with Fiber and Disk Laser*, International Congress on Applications of Lasers & Electro-Optics, 2006/1, 404, https://doi.org/10.2351/1.5060827.

[14]  LEVICHEV N., RODRIGUES G.C., GARCIA A.T., DUFLOU J.R., 2020, *Trim-Cut Technique for Analysis of Melt Flow Dynamics in Industrial Laser Cutting Machine*, Procedia CIRP, 95, 858–863, https://doi.org/10.1016/j.procir.2020.01.15.

[15]  AHN D.G., YOO Y.T., 2006, *Investigation of Cutting Characteristics in the Sharp Edge for the Case of Cutting of a Low Carbon Steel Sheet Using High-Power CW Nd:YAG Laser*, Journal of KWS, 24/4, Available: https://koreascience.kr/main, Accessed 9 July 2024.

[16]  MCDONNELL M.D.T., ARNALDO D., PELLETIER E., GRANT-JACOB J.A., PRAEGER M., KARNAKIS D., EASON R.W., MILLS B., 2021, *Machine Learning for Multi-Dimensional Optimisation and Predictive Visualisation of Laser Machining*, Journal of Intelligent Manufacturing, 32/5, 1471–1483, https://doi.org/10.1007/s10845-020-01717-4.

[17]  WANDERA, C.; KUJANPAA, V.; SALMINEN, 2011, *A. Laser Power Requirement for Cutting Thick-Section Steel and Effects of Processing Parameters on Mild Steel Cut Quality*, Proceedings of the Institution of Mechanical Engineers, Part B J. Eng. Manuf., 225, 651–661, https://doi.org/10.1177/09544054JEM1971.

[18]  TEIXIDOR, D., CIURANA, J., RODRIGUEZ, C.A., 2014, *Dross Formation and Process Parameters Analysis of Fibre Laser Cutting of Stainless-Steel Thin Sheets*, International Journal of Advanced Manufacturing Technology, 71/9–12, 1611–1621, https://doi.org/10.1007/s00170-013-5599-0.

[19]  WANG, X. N., SUN, Q., ZHENG, Z., DI, H.S., 2017, *Microstructure and Fracture Behavior of Laser Welded Joints of DP Steels with Different Heat Inputs*, Materials Science and Engineering: A, 699, 18–25, https://doi.org/10.1016/j.msea.2017.05.078.

[20]  GHANI R., 2021, *Integration of FMECA and Statistical Analysis for Predictive Maintenance*, Journal of Applied Research in Technology & Engineering, 2/1, 33–37, https://doi.org/10.4995/jarte.2021.14737.

[21]  CATIC D., JEREMIC B., DJORDJEVIC Z., MILORADOVIC N., 2011, *Criticality Analysis of the Elements of the Light Commercial Vehicle Steering Tie-Rod Joint*, Strojniski vestnik – Journal of Mechanical Engineering, 57/6, 495–502, http://dx.doi.org/10.5545/sv-jme.2010.077.